\pdfoutput=1
\documentclass[a4paper,superscriptaddress,prl,twocolumn,amsmath,showpacs]{revtex4-1}
\usepackage{graphicx}
\usepackage[update,prepend,prefersuffix=true]{epstopdf}

\begin{document}


\title{A minimally constrained model of self-organized helical states in reversed-field pinches} 



\author{G.R. Dennis}
\email[Electronic address: ]{graham.dennis@anu.edu.au}
\affiliation{Research School of Physics, The Australian National University}
\author{S.R. Hudson}
\affiliation{Princeton Plasma Physics Laboratory, Princeton University}
\author{D. Terranova}
\author{P. Franz}
\affiliation{Consorzio RFX, Associazione Euratom-ENEA sulla Fusione, corso Stati Uniti 4, 35127 Padova, Italy}
\author{R.L. Dewar}
\affiliation{Research School of Physics, The Australian National University}
\author{M.J. Hole}
\affiliation{Research School of Physics, The Australian National University}


\date{\today}

\begin{abstract}
We show that the self-organized single-helical-axis (SHAx) and double-helical-axis (DAx) states in reversed field pinches can be reproduced in a minimally constrained equilibrium model using only five parameters.  This is a significant reduction on previous representations of the SHAx which have required an infinite number of constraints.  The DAx state, which has a non-trivial topology, has not previously been reproduced using an equilibrium model that preserves this topological structure.  We show that both states are a consequence of transport barrier formation in the plasma core, in agreement with experimental results.  We take the limit of zero pressure in this work, although the model is also valid for finite pressure.
\end{abstract}


\maketitle 

A major goal of the theory of complex physical systems is to find relatively simple organizing principles that operate when systems are strongly driven.  A famous early example of such a universal principle is the Taylor relaxation principle \citep{Taylor:1986}, which postulates that a plasma tends to minimize its total magnetic energy subject only to the constraints of conservation of global magnetic flux and global magnetic helicity.  This principle has been successful in describing the classical behavior of the core region of Reversed Field Pinch (RFP) experiments where many magnetohydrodynamic (MHD) modes resonate on different plasma layers.  These modes form overlapping magnetic islands and result in a chaotic field region extending over most of the plasma volume \citep{DAngelo:1996}.  The consequent destruction of magnetic surfaces leads to modest confinement in this regime, and was thought to prevent fusion power development with the RFP.

This classical paradigm of the RFP as a chaotic plasma with modest confinement properties has been challenged in recent years with the observation of the high-confinement quasi-single-helicity (QSH) regime \citep{Escande:2000,Martin:2003}.
The transition to the QSH regime occurs as the plasma current is increased ($>$1MA), and a single dominant helical mode arises spontaneously.  A second (helical) magnetic axis forms associated with this helical mode and this state is known as the double-axis state (DAx) \citep{Puiatti:2009}.  As the current is increased further a topological change in this magnetic configuration is observed: the main magnetic axis disappears in a saddle-node bifurcation \citep{Escande:2000a}, forming a helical plasma column despite the axisymmetric plasma boundary.  This is the single-helical-axis (SHAx) state \citep{Escande:2000a} which has recently been observed in RFX-mod \citep{Lorenzini:2008,Lorenzini:2009} and is associated with strong electron transport barriers and significantly improved plasma confinement.  

As the DAx and SHAx states are formed by a self-organized process, they should be describable in terms of a small number of parameters.  Taylor's theory was successful in describing the classical chaotic regime in the core of the RFP with only two parameters, however it is unable to describe the self-organized states in the QSH regime because, although it has a helical solution for sufficiently high magnetic helicity \citep{Taylor:1986}, the helical pitch of this solution is opposite to that of the observed QSH states \citep{Escande:2000}.

\begin{figure}
  \includegraphics[width=9cm]{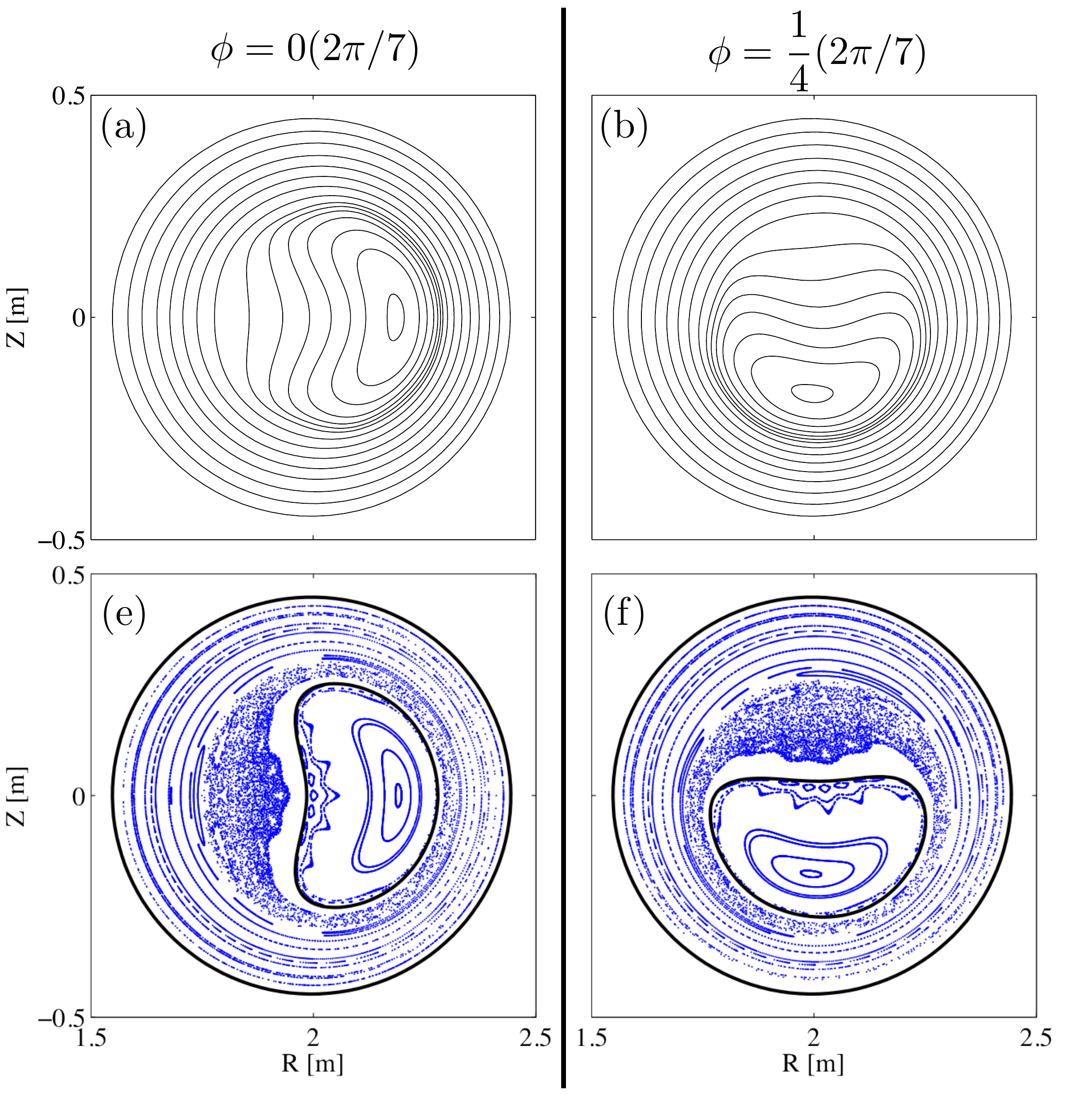}
  \caption{\label{fig:IdealMHDFluxSurfaces}Comparison of the ideal MHD representation of the SHAx state in RFX-mod and the minimal model (MRXMHD) of this state presented in this work.  Figures (a)--(b) show the (poloidal) magnetic flux contours of the ideal MHD plasma equilibrium at toroidal angles covering one quarter of the period of the helical solution.  Figures (c)--(d) show Poincar\'e plots of the minimal model at the same toroidal locations as (a)--(b).  The thick black lines mark the location of the transport barrier separating the two plasma volumes.  The minimal model corresponds to the $\lambda=0.3$ configuration of Figure~\ref{fig:EnergyPlotWithAxisymmetric}.}
\end{figure}

The SHAx state in the QSH regime has been reconstructed using the ideal MHD equilibrium framework assuming continuously nested magnetic flux surfaces \citep{Terranova:2010} (see Figure~\ref{fig:IdealMHDFluxSurfaces}(a)--(b)).  The continuously nested flux surface assumption typically used with ideal MHD requires the specification of the enclosed toroidal and poloidal fluxes as a function of the magnetic flux surface.  These continuous flux functions are an infinite number of constraints on the plasma equilibrium, and are therefore not a natural description of the self-organized QSH regime.  The continuously nested flux surface assumption also prevents the description of non-trivial magnetic structure such as islands and chaotic regions.  These constraints prevent this equilibrium framework from describing the DAx state, which has two magnetic axes.  This Letter presents the results of a generalization of Taylor's theory which describes both the SHAx and DAx states in the QSH regime with a minimum number of free parameters.  Both states are naturally reproduced as a result of a single transport barrier in the core of the plasma. This is in agreement with experimental observations of an electron transport barrier surrounding the core of the plasma in the SHAx state \citep{Lorenzini:2009}.

\begin{figure}
  \includegraphics[width=5cm]{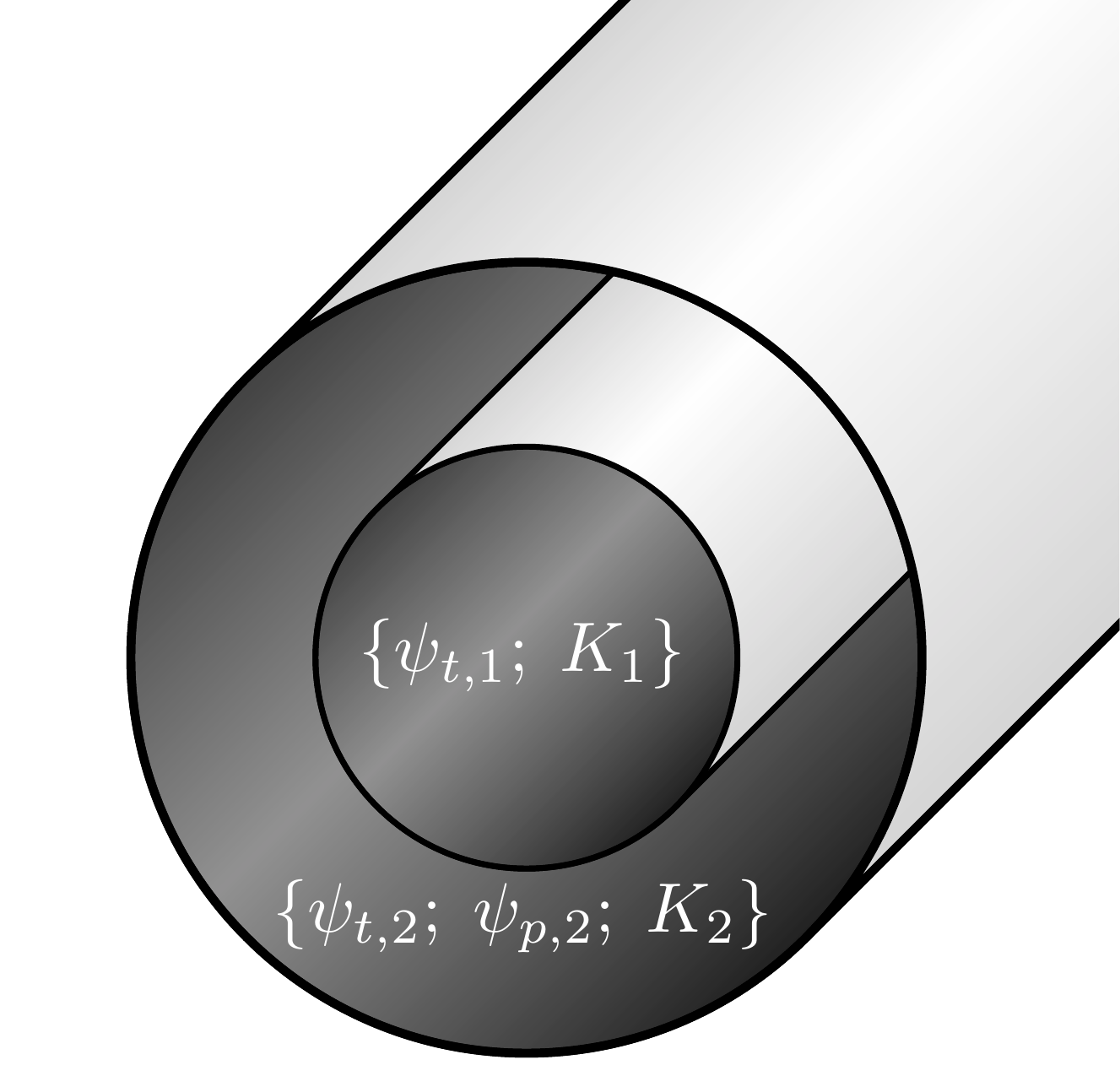}
  \caption{\label{fig:MRXMHDConceptual}Five constraints are needed to specify the two-volume MRXMHD plasma equilibrium: the toroidal flux in each volume, $\psi_{t,i}$; the poloidal flux in the outer volume, $\psi_{p,i}$; and the magnetic helicities in each volume, $K_i$.}  
\end{figure}

A stable plasma equilibrium is a constrained minimum of the plasma energy
\begin{align}
  W &= \int \left(\frac{\mathbf{B}^2}{2\mu_0} + \frac{p}{\gamma - 1} \right)\, d^3 x, \label{eq:PlasmaEnergy}
\end{align}
where $\mathbf{B}$ is the magnetic field, $\mu_0$ is the permeability of free space, $p$ is the plasma pressure and $\gamma$ is the ratio of specific heats.  The plasma states over which $W$ is minimized must be constrained to avoid the trivial $\mathbf{B}=0$ solution.  The traditional approach of ideal MHD is to consider only states with nested magnetic flux surfaces with the enclosed toroidal and poloidal fluxes specified as a function of the magnetic flux surface.  
This Letter considers a wider class of plasma equilibria by relaxing the \emph{continuously} specified constraints of the traditional equilibrium framework to a finite number of \emph{discrete} constraints.  We apply the MRXMHD framework \citep{Hole:2007,Hudson:2007}, which is a generalization of Taylor's relaxation theory, in which the plasma is partitioned into a finite number of nested regions $\mathcal{R}_i$ that independently undergo Taylor-relaxation.  The plasma regions are separated by ideal transport barriers $\mathcal{I}_i$ that are also assumed to be magnetic flux surfaces  (the two-volume case is illustrated in Figure~\ref{fig:MRXMHDConceptual}).  In the MRXMHD framework plasma equilibria are obtained by minimizing \eqref{eq:PlasmaEnergy} subject to discrete constraints on the enclosed magnetic fluxes, magnetic helicity and thermodynamic quantities in each plasma region.  As part of the energy minimization process the geometry of the ideal transport barriers are varied to ensure that force-balance is achieved across each barrier.

As the QSH regime is a high-current regime the effect of pressure can be negligible, and this is the case for the configurations considered here.  The limit of zero pressure has been taken in the ideal MHD equilibrium presented in Figure~\ref{fig:IdealMHDFluxSurfaces}(a)--(b) and will be assumed in the remainder of this Letter.

The magnetic helicity constraint in MRXMHD and Taylor's relaxation theory is a topological constraint related to the Gauss linking number of flux tubes and is the most preserved of the ideal MHD invariants in the presence of small amounts of resistivity \citep{Taylor:1986,Hameiri:1982,Qin:2012}.  Taylor's relaxation theory preserves the magnetic helicity globally throughout the entire plasma and can be physically interpreted as the idea that a weakly resistive plasma will evolve to minimize the plasma energy, but the magnetic field cannot untangle itself.  The MRXMHD framework extends this idea to include a number of transport barriers that partition the plasma and prevent complete reconnection.  In the MRXMHD framework the magnetic topology within each plasma region is completely free; only the ideal transport barriers are constrained to be magnetic flux surfaces.

In this Letter we seek to use the MRXMHD model to develop a minimal model of the RFP QSH regime.  The smallest number of constraints in the MRXMHD model is when the entire plasma is taken as a single volume without any transport barriers partitioning it.  This is exactly Taylor's relaxation theory, which we already know is insufficient to model the observed QSH states.  In the opposite limit of an infinite number of interfaces, MRXMHD approaches ideal MHD \citep{Dennis:2013a}, and as ideal MHD can describe the SHAx state (but not the DAx state) we can expect that MRXMHD will be able to reproduce this state given a sufficiently large number of interfaces.  As we seek a minimal model of both the SHAx and DAx states, the next simplest possible model is that with two plasma volumes separated by a transport barrier.

The MRXMHD model with two volumes requires the specification of five constraints (see Figure~\ref{fig:MRXMHDConceptual}): the toroidal fluxes and the helicity in each volume, and the poloidal flux in the outer volume \footnote{The poloidal flux is not constrained in the inner volume because \emph{a priori} there is no magnetic axis for it to be defined with respect to.  The same situation occurs in Taylor's relaxation theory in which the global toroidal flux is constrained, but the global poloidal flux is not.  In the outer plasma region the poloidal flux can be defined relative to the transport barrier.}.  A scheme is needed for obtaining values for these quantities that are appropriate for the QSH regime.  We do this by taking an ideal MHD equilibrium for the SHAx state \footnote{The ideal MHD state used is a representation of the SHAx state in RFX-mod obtained by \citet{Terranova:2010} (Figure~2 of \citep{Terranova:2010}).}, choosing one of the flux surfaces (labelled by $\lambda$) to act as the transport barrier in our model, and then computing the values of the constraints in each plasma region.  As there is freedom in which flux surface $\lambda$ of the ideal MHD equilibrium to choose to act as the transport barrier, this procedure defines a one-dimensional line $(\psi_{t,1}(\lambda), K_1(\lambda), \psi_{t,2}(\lambda), \psi_{p,2}(\lambda), K_2(\lambda))$ of the five-dimensional $(\psi_{t,1}, K_1, \psi_{t,2}, \psi_{p,2}, K_2)$ parameter space.  The one-dimensional line in the parameter space has been chosen to be consistent with the SHAx state.  The values of these five parameters are then used as inputs to our MRXMHD model, and the geometry of the $\lambda$ ideal MHD flux surface is taken as a convenient initial guess for the geometry of the transport barrier in our model.  The geometry of the transport barrier is necessarily varied as part of the energy minimization process to obtain the plasma equilibrium.  We use the Stepped Pressure Equilibrium Code \citep{Hudson:2012} to compute the MRXMHD solutions presented in this Letter.

\begin{figure}
  \includegraphics[width=8cm]{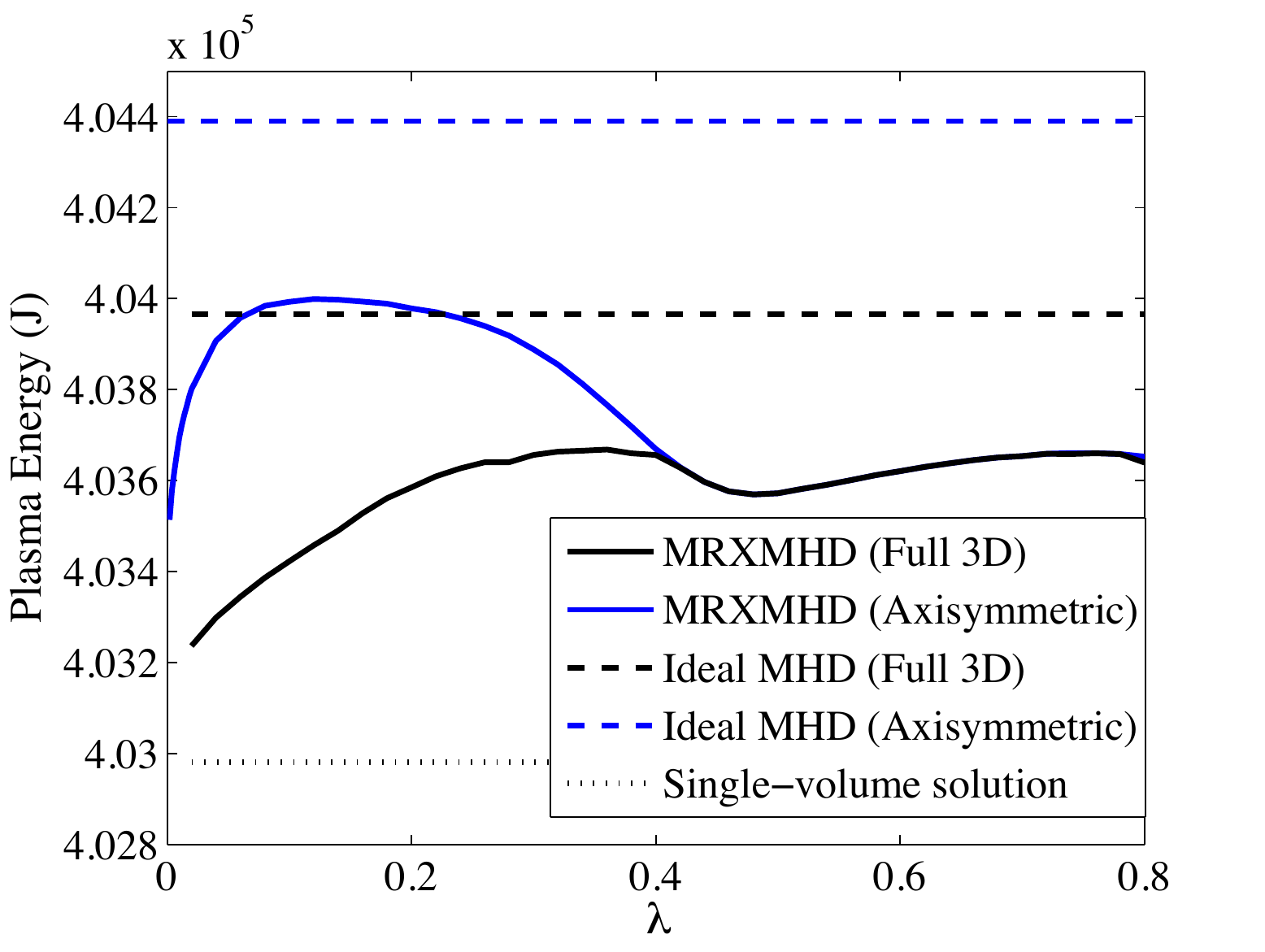}
  \caption{\label{fig:EnergyPlotWithAxisymmetric}  Plot of the plasma energy for different plasma equilibria as a function of the magnetic flux surface $\lambda$ chosen as the transport barrier in the MRXMHD model.  The ideal MHD flux surfaces are labelled by the normalized enclosed poloidal flux ($0 \leq \lambda \leq 1$).
  }
\end{figure}

The results of the one-dimensional parameter scan over $\lambda$ are plotted in Figure~\ref{fig:EnergyPlotWithAxisymmetric}, which depicts the minimum energy for each set of constraints and compares them to the single volume case (no transport barrier; Taylor's relaxation theory) and the continuously nested flux surface case (ideal MHD).  Also plotted in blue are the minimum energies found for the same constraints but restricting the minimization procedure to only consider axisymmetric plasmas.  The plasma energies obtained for different constraints (different values of $\lambda$) cannot be directly compared; this would be akin to comparing the energies of ideal MHD equilibria with different safety-factor profiles.  A comparison can only be made between solutions with consistent constraints, i.e.\ solutions with the same value of $\lambda$.  A partial order is expected of the solutions plotted in Figure~\ref{fig:EnergyPlotWithAxisymmetric}: 
\begin{enumerate}
  \item The energy of the axisymmetric solution for a given equilibrium model should be an upper bound for the energy of the solution with full 3D freedom allowed as a wider class of plasma states are considered in the latter case.  When the energies of the axisymmetric and full 3D solutions are different the solution with full 3D freedom must have non-axisymmetric structure.
  \item The single-volume equilibrium has fewer constraints than the two-volume MRXMHD 3D solution or the ideal MHD 3D solution and is therefore a lower bound for the energies of these solutions.  As the single-volume solution is also axisymmetric, it is also a lower bound for the energies of the axisymmetric MRXMHD and ideal MHD solutions.
  \item As the ideal MHD equilibrium has more constraints than the two-volume MRXMHD solution, it is an upper bound for the energies of those solutions with the same amount of geometric freedom, for example, the axisymmetric ideal MHD solution is an upper bound for the energy of the axisymmetric two-volume MRXMHD solution.
\end{enumerate}
This expected partial ordering is borne out in Figure~\ref{fig:EnergyPlotWithAxisymmetric}. 


In Figure~\ref{fig:EnergyPlotWithAxisymmetric}, for $\lambda \lesssim 0.4$ the differences between the energies of the single transport barrier solutions and the corresponding solutions with assumed axisymmetry indicate that a non-axisymmetric solution develops associated with a transport barrier in the core region.  This non-axisymmetric structure is helical in nature as shown in the Poincar\'e plots in Figure~\ref{fig:IdealMHDFluxSurfaces}(c)--(d), which have the same qualitative structure as the ideal MHD solution with continuous flux surfaces in Figure~\ref{fig:IdealMHDFluxSurfaces}(a)--(b) with the exception of additional topological structure such as islands and chaotic regions that cannot be represented in the ideal MHD solution.  The similarity between these two figures demonstrates that only a single transport barrier is required to reproduce the self-organized SHAx state.  This is the first time such a nontrivial magnetic topology has been reproduced non-perturbatively within a plasma equilibrium description.

The energy differences between the solutions plotted in Figure~\ref{fig:EnergyPlotWithAxisymmetric} are very small, at most about $0.3\%$ of the total plasma energy.  A similar situation of a small energy difference between identically constrained axisymmetric and helical equilibria was previously observed by \citet{Cooper:2010a} in the context of tokamak plasmas.  \citeauthor{Cooper:2010a}\ argued that their slight energy difference suggested that transitions between their axisymmetric and helical states could occur easily.  In the present context, while the SHAx state is observed to spontaneously collapse to an axisymmetric configuration \citep{Lorenzini:2009}, we would argue that a comparison of the absolute plasma energies necessarily presupposes that zero-energy is a relevant baseline for comparison.  A better reference energy would be that of a minimally constrained plasma, i.e.\ the energy of the single-volume Taylor relaxed state.  Seen in this light, the energy difference between the axisymmetric and full 3D ideal MHD solutions in Figure~\ref{fig:EnergyPlotWithAxisymmetric} is about 30\% of the maximum amount of energy that the axisymmetric configuration could lose while still remaining a plasma.

\begin{figure*}
  \includegraphics[width=18cm]{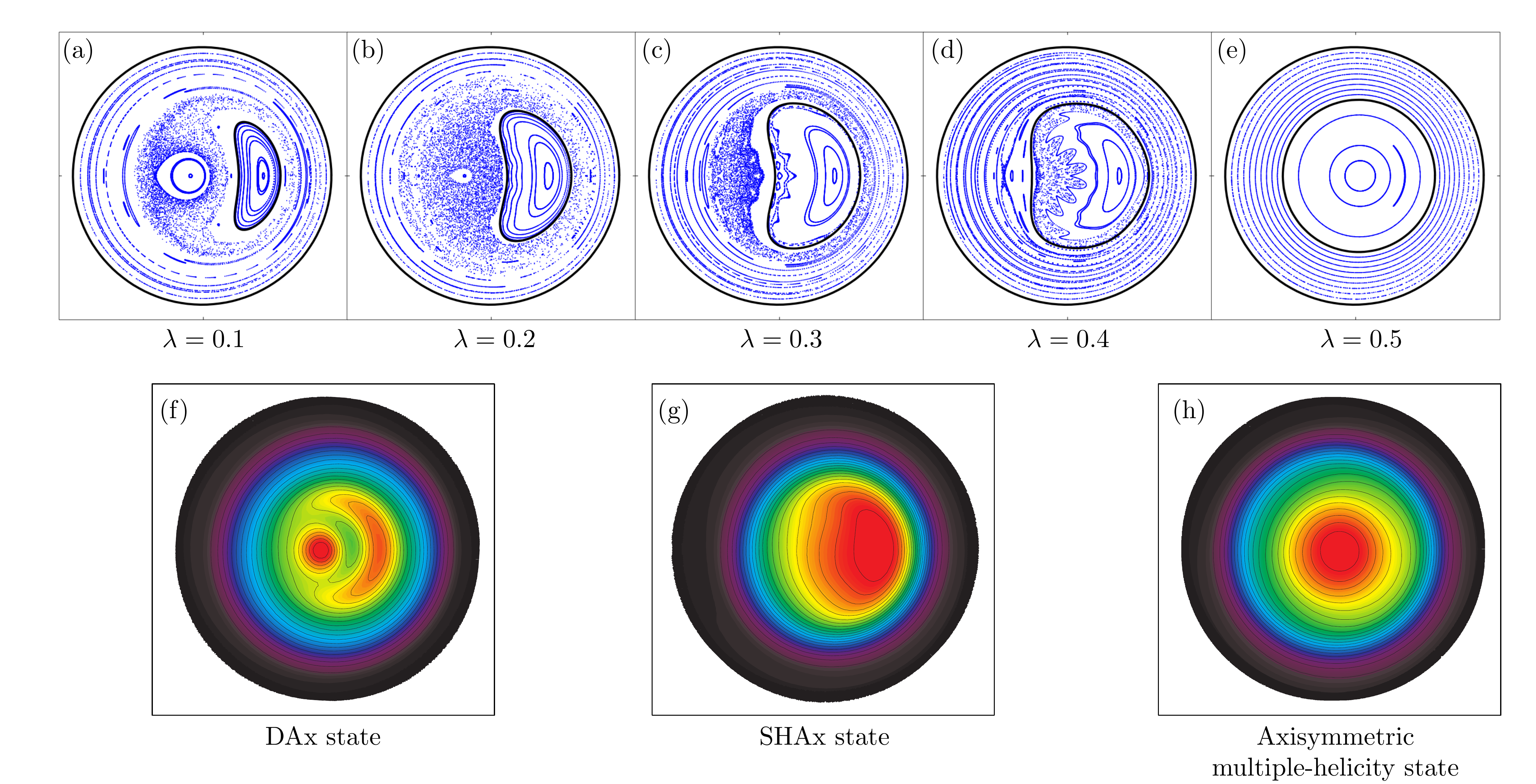}
  \caption{\label{fig:PoincareProgression} A comparison between the magnetic structure computed from the model presented in this Letter and soft X-ray measurements from RFX-mod.  Figures (a)--(e): Poincar\'e plots for single-barrier MRXMHD equilibria for different values of $\lambda$, the ideal MHD flux surface chosen to act as the transport barrier.  Figures (f)--(h): Tomographic inversions of soft X-ray emissivity for plasmas in the  (f) Double-Axis state, (g) the Single Helical-Axis state and (h) the axisymmetric multiple-helicity state.  Note that the structures in figures (f) and (g) are helical in nature.}
\end{figure*}

Figure~\ref{fig:PoincareProgression} illustrates Poincar\'e plots for a range of values of $\lambda$ and compares these to a tomographic inversion of soft X-ray measurements \footnote{To ensure that DAx-like states would be correctly reproduced by the tomographic inversion algorithm, no particular topology was assumed, i.e.\ the reconstructions pictured in Figure~\ref{fig:PoincareProgression}(f)--(h) are the best emissivity distributions that reproduce the experimental measurements, independently of the assumed magnetic flux surface topology.} (which are a proxy for temperature) from RFX-mod.  This figure demonstrates that we can reproduce DAx-like solutions in (a) and (b) as well as the SHAx-like solutions in (c) and (d) (see also Figure~\ref{fig:IdealMHDFluxSurfaces}).  The DAx-like solutions are in qualitative agreement with the soft X-ray measurements presented in (f), as well as reconstructed Poincar\'e plots from the MST device \citep{Martin:2003} and RFX-mod \citep{Martin:2009}.  As $\lambda$ increases and the transport barrier leaves the plasma core and approaches the edge, the solution becomes mostly axisymmetric for $\lambda \gtrsim 0.5$ [see Figure~\ref{fig:PoincareProgression}(e)], resembling the classical multiple-helicity regime depicted in Figure~\ref{fig:PoincareProgression}(h).  This suggests that the QSH regime is correlated to the formation of a transport barrier near the plasma core.  The existence of a transport barrier near the plasma core is supported by experimental measurements in RFX-mod \citep{Lorenzini:2009}. 

This Letter has demonstrated a minimal model that is able to qualitatively reproduce the magnetic structure of both the self-organized SHAx and DAx states in the QSH regime of RFPs.  Previous recreations of the SHAx state have required an infinite number of constraints to parameterize the model; the model presented in this Letter has only five: the enclosed toroidal fluxes and helicities in the inner and outer volumes, and the enclosed poloidal flux in the outer volume. Fewer constraints are not possible as there is no MRXMHD model with 3 or 4 constraints, and Taylor's relaxation theory, which has 2 constraints, cannot reproduce the QSH regime of RFPs.

\begin{acknowledgments}
The authors gratefully acknowledge support of the U.S. Department of Energy and the Australian Research Council, through Grants DP0452728, FT0991899, and DP110102881.  We acknowledge the use of VMEC \citep{Hirshman:1986} from S.P.~Hirshman, and we thank D.~Escande for his helpful comments.
  
\end{acknowledgments}

\bibliography{rfp_paper.bib}

\end{document}